# Experimental Evaluation of a Checklist-Based Inspection Technique to Verify the Compliance of Software Systems with the Brazilian General Data Protection Law

Diego André Cerqueira
Systems Engineering and Computer Science Graduate Program – Federal University of Rio de Janeiro, Brazil
https://orcid.org/0000-0003-4861-3394

Rafael Maiani de Mello
Computing Institute – Federal University of Rio de Janeiro, Brazil
https://orcid.org/0000-0002-9877-3946

Guilherme Horta Travassos
Systems Engineering and Computer Science Graduate Program – Federal University of Rio de Janeiro, Brazil
https://orcid.org/0000-0002-4258-0424

*Abstract*—Background/Context: Recent laws to ensure the security and protection of personal data establish new software requirements. Consequently, new technologies are needed to guarantee software quality under the perception of privacy and protection of personal data. Therefore, we created a checklist-based inspection technique (LGPDCheck) to support the identification of defects in software artifacts based on the principles established by the Brazilian General Data Protection Law (LGPD). Objective/Aim: To evaluate the effectiveness and efficiency of LGPDCheck for verifying privacy and data protection (PDP) in software artifacts compared to ad-hoc techniques. Method: To assess LGPDCheck and ad-hoc techniques experimentally through a quasi-experiment (two factors, five treatments). The data will be collected from IoT-based health software systems built by software engineering students from the Federal University of Rio de Janeiro. The data analyses will compare results from ad-hoc and LGPDCheck inspections, the participant's effectiveness and efficiency in each trial, defects' variance and standard deviation, and time spent with the reviews. The data will be screened for outliers, and normality and homoscedasticity will be verified using the Shapiro-Wilk and Levene tests. Nonparametric or parametric tests, such as the Wilcoxon or Student's t-tests, will be applied as appropriate.

## I. INTRODUCTION

The massive production of data caused by the popularization of mobile devices, intelligent devices, and the use of social networks in recent decades has constantly challenged the right to privacy and protection of personal data in the digital age [1]. In 2016, the General Data Protection Regulation (GDPR) was approved to keep up with these challenges. Then, in 2018, the legal system in Brazil shifted towards the subject of Privacy and Data Protection (PDP) through the approval of the Brazilian General Data Protection Law (LGPD) [2].

Although GDPR and LGPD do not directly address the compliance of software products with PDP regulations, we can observe that these regulations impact several activities from the software development cycle [3]. Companies need to conceive and tailor their software products and processes to meet the legal requirements outlined in these laws. However, software professionals lack specialized support to reach this goal [4].

Considering the regulation imposed by the LGPD in Brazil, we need to ensure the compliance of software systems under construction and existing ones with the LGPD principles, avoiding possible penalties for organizations. In this sense, we observe a persistent gap in software systems implementing the rights, duties, and responsibilities required by the LGPD principals. One possible reason for this gap is the complex and specific legal language composing the description of LGPD, uncommon for software professionals. Besides, due to the nature of LGPD, its content does not provide concrete and objectively verifiable situations in any domain, including software systems. Thus, the software engineering community should bridge this gap by providing evidence-based technologies to support the integration of LGPD and other PDP regulations into the software development practice [5].

Despite this gap, the application of the LGPD is already part of the daily life of several Brazilian teams. Besides, the use of LGPD in software development has been investigated. In the organizational context, the LGPD4BP (LGPD for Business Process) [6] technique aims to assess the compliance of business processes with the LGPD. LGPD4BP checklist and the process were conceived based on the subject's rights and sections from LGPD. Mendes et al. [7] proposed the initial version of a checklist to verify the compliance of software systems with some attributes from LGPD and GDPR. This checklist was then extended by [8] to support verification of Internet of Things (IoT) based systems. Despite the relevance of the technique and the attributes treated in these technologies, we could not identify any supporting content (such as guidelines) for practitioners to accurately interpret and apply the LGPD and PDP principles in the specific context of software development. Besides, the techniques above aim to support developers in verifying already existing software systems, not covering the building of new ones. In this sense, it is important to note that the investigation presented in [3] revealed that development teams perceive a high impact of addressing LGPD on requirements specification and other software development activities.

Besides, we could not identify checklists for verifying the compliance of software systems with other PDP laws.

Based on the previously mentioned investigations, we identified the opportunity of conceiving an evidence-based software technology to support software professionals in verifying the compliance of software artifacts with LGPD principles by guiding the community of practice to interpret these principles in the context of software development and its related activities [5]. In this way, the ongoing research presented in this article aims to answer the following question: *"Can a checklist-based inspection technique improve the effectiveness and efficiency of inspections for verifying the compliance of software systems with the Brazilian General Data Protection Law (LGPD) principles?"*

Therefore, this work aims to evaluate a checklist-based inspection technique (*LGPDCheck*) [20] to support the identification of defects in software artifacts produced in the different stages of the software development cycle from the perspective of the restrictions and conditions established by the LGPD principles. To achieve this goal, we understand that *LGPDCheck* should offer, in addition to the checklist, supporting content to guide inspectors to contextualize the LGPD principles to the software development and evolution practice.

The development of *LGPDCheck* follows a research methodology inspired by Design Science [9]. The objective of Design Science is to solve real-world problems by instrumentalizing specific knowledge of a field or area and, thus, help specialists and professionals build solutions based on experiences in the field [10]. In this way, we planned the following five research steps, from which we already performed the four first ones: (1) conducting interviews with specialists in verifying the compliance of software systems with the LGPD; (2) bibliographic review of PDP application in software systems; (3) LGPD content analysis; (4) construction of the artifacts comprising *LGPDCheck*; (5) experimental evaluation of *LGPDCheck*.

In this registered report, we present the design of the confirmatory study we planned to evaluate *LGPDCheck*. This study aims at characterizing the feasibility of the inspection technique in terms of effectiveness and efficiency. For this purpose, we recruited three software development teams working in a real IoT-based system. We also introduce in this report the inspection technique and discuss related work.

## II. BACKGROUND

### A. Privacy and data protection laws

With the approval of GDPR in Europe in 2016, countries like Brazil were positively influenced to push for privacy and data protection regulations. Thus, the Brazilian General Data Protection Law (LGPD) - 13,709/2018 was approved in 2018. The LGPD presents essential privacy and data protection definitions to Brazilian society. In addition, the law defines the main concepts of PDP, such as personal data, processing agents (controller and processor), data processing, consent, and data subject [2].

The Brazilian law is grounded on ten principles that are the basis for any personal data processing. They are *purpose, adequacy, necessity, free access, data quality, transparency, security, prevention, non-discrimination, and accountability* [2]. It is worth mentioning that both the European (GDPR) and the Brazilian (LGPD) regulations present principles for processing and/or handling personal data from subjects. However, these regulations are, by conception, principle-based. Thus, the LGPD does not show how its principles should be concretely addressed in different contexts where personal data is processed, including the development or evolution of software systems. In this way, it is important to note that implementing PDP practices continues to challenge the software community. In [4], the authors highlight the difficulties caused by the ambiguity or imprecision of data protection laws in presenting its principles and other elements. In addition, the authors highlight the challenge of identifying how to comply with a given principle. Finally, Hadar et al. [11] highlight the risk of conceptual confusion between the concepts of PDP and information security, followed by the need for methodologies that support the implementation of both concepts in the different stages of the software life cycle [12].

In the context of PDP, one may see that the concept of Privacy by Design (PbD) is widespread in the field. PbD aims to establish seven fundamental principles that guide the implementation of privacy since the first steps of product development. From physical to digital products, the PbD framework of principles proposes prioritizing privacy, making it the primary objective to achieve [13]. For this purpose, Shapiro [5] suggests the design of technologies that offer (1) an adequate translation of abstract privacy principles, privacy risk models, and privacy mechanisms into implementable requirements, (2) the integration of these requirements into an appropriate process; and; (3) incorporation of the process into the development lifecycle.

### B. Software Inspection

Software inspection means visually verifying a software artifact to identify defects, including anomalies, faults, errors, and deviations from standards and specifications [14]. Software inspection is considered a low-cost technology that promotes the quality of software artifacts since the early stages of the development process, ranging from requirements [15][17] to testing stages [18]. Consequently, software inspections increase software professional productivity, reduce rework costs [15], and allow them to detect and fix defects that would be neglected in the various stages of the software development and evolution process [16].

One of the decisive factors in the planning of software inspections is the selection of the inspection technique [19]. As an alternative to ad-hoc inspections, several inspection techniques are available in the technical literature, including checklist-based ones and reading techniques [20]. Checklist-based inspection techniques are composed of questions with "Yes/No" answers, leading the inspector to reflect on different quality perspectives while inspecting a particular software artifact [21]. In addition, some checklists-based inspection techniques allow the calibration of their checklists according to the scope of each inspection.

The authors of this registered report are members of the Experimental Software Engineering Research Group (ESE).

In the last decades, the ESE group[1] members have been involved in several research projects for building inspection techniques for different software artifacts (Table I). The feasibility of these techniques is evaluated through confirmatory studies, typically quasi-experiments with students. In these studies, the inspection technique's *effectiveness* and *efficiency* are measured and compared with *ad-hoc* inspections. In some cases, we also evaluated the acceptance of the technology. The composition of the experimental design varies according to the samples and the software artifacts available.

TABLE I. INSPECTION TECHNIQUES BUILT BY THE ESE GROUP.

| Name | Ref. | Scope |
|------|------|-------|
| OORTs | [23] | a family of reading techniques to support the verification and validation of different UML-based object-oriented diagrams. |
| OO-PBR | [24] | perspective-based reading technique for identifying defects in requirements documents written in natural language. |
| ArqCheck | [25] | checklist-based inspection technique for detecting defects in architectural documents. |
| ActCheck | [26] | checklist-based inspection technique focused on semantic verification of activity diagrams describing requirements specifications. |
| BPCheck | [27] | checklist-based inspection technique to detect defects in BPMN models. |
| FMCheck | [22] | checklist-based inspection technique to support the detection of defects in feature models describing Software Product Lines |
| ScenarIoTCheck | [28] | checklist-based inspection technique supporting the quality of the description of IoT (Internet of Things) scenarios. |

## III. LGPDCHECK

We understand that an inspection technique is a promising strategy for supporting professionals in verifying the compliance of software artifacts with LGDP principles. However, to build this technique, we first must understand how the LGPD principles have been verified in software systems. In this sense, we interviewed two LGPD specialists to characterize their practice in verifying the compliance of software systems with LGPD principles. These specialists are lawyers with a specialization in Digital Law. Besides, they have practical experience in law and technology, ranging from four to six years. This experience includes applying their legal background, including LGPD, in the software life cycle.

We conducted both interviews remotely, each one lasting approximately 60 minutes. Through these interviews, it was possible to characterize, anecdotally, how the specialists verify the compliance of software systems with LGPD principles and PDP practices. Furthermore, these findings revealed relevant concerns of the specialists in their verifications and practices for specifying software systems in compliance with LGPD. The problems raised by the specialists include the following:

- An attempt to implement the principles of PDP from the design of the system's functionalities and;
- The absence of guides for implementing the LGPD in software development.

The practices include:

- The adoption of *use cases* for specifying the data to be manipulated in implementing system functionalities;
- The use of additional structures called *data maps* to record relevant information, such as the data origin and the data type;
- The heavy use of their legal background and expertise in LGPD during feature brainstorming sessions and daily meetings with the software development team;
- Efforts to create *"Privacy by Design"* products in attempts to implement PDP from feature design;
- Data mapping flow: This artifact records which personal data and functionalities use them and;
- Reverse engineering of existing systems: The expert asks a set of questions during the reverse engineering process, such as "*What data is used?*", "*Why do I need them for?*"

The knowledge gathered through the interviews revealed the need for the inspection technique to support verifying both running software systems and requirements specifications. Besides, we identified the need to develop a support material associating the generic LGPD principles with desired behaviors and corresponding defects in software systems. Considering these directions, we opted to create a checklist-based inspection technique to support the verification of the LGPD principles once the verification items of a checklist do not aim to determine specific defects but lead inspectors to reflect on their incidence [19].

We built the first version of *LGPDCheck*, a checklist-based inspection technique designed to support the identification of defects (*omission, incorrect fact, ambiguity, inconsistency,* and *extraneous information*) in software systems addressing their compliance with the LGPD principles [20]. We identified that verifying certain LGPD principles may require gathering information from requirements specifications and running systems. Besides, addressing LGPD principles may also lead organizations to establish rules that their systems should follow. Thus, we built *LGPDCheck* to support multi-level inspections. This multi-level inspection aims to cover the verification of the following artifacts:

- *Level I – System Requirements*: The verification occurs in the requirements document of the software systems to be built or under construction. The goal is to verify whether the description of the system functionalities and expected data processing comply with the PDP principles.
- *Level II – Systems in execution*: The verification takes place at the functionality level during the execution of the software system. This level aims to verify whether the software system provides services and functionalities in compliance with the principles of PDP.

---

[1] http://lens-ese.cos.ufrj.br/ese/publicacoes-do-grupo-ese/

- *Level III – Organizational Requirements*: The verification occurs at the organizational level. It aims to identify corporate rules, procedures, and guidelines that define how the company systems should implement PDP rules.

We considered the ten principles of LGPD and their definitions to propose these three levels of inspection mentioned above. By focusing on an inspection level, we expect that inspectors can optimize their assessment efforts, applying the most relevant verification items. The relationship between the LGPD principles was considered from their definitions of the law itself [2]. The principles prioritized for Level I (system requirements) are *purpose, adequacy*, and *necessity*. The principles prioritized for Level II (Running systems) are *free access, quality of the data,* and *transparency*. Finally, security, *prevention, non-discrimination,* and *accountability* are prioritized to Level III (Organizational Practices).

*A. LGPDCheck Resources*

As with any other checklist-based inspection technique, *LGPDCheck* has verification items, i.e., questions that help inspectors identify possible defects. In addition, however, the technique also provides a *supporting material* composed of ten *frames*, each introducing each LGPD principle from the perspective of software systems. The supporting material also includes a glossary of LGPD terms adapted to the context of software systems.

Each frame comprises the LGPD principle, its interpretation for software systems, examples of violating the principle in software systems, and expected actions from the system owners when designing the system's data processing activities to meet the corresponding principle.

Table II presents one of the *LGPDCheck* frames from the supporting material, addressing the principle of *necessity*. The end of each frame provides the *verification items* addressing each principle. For each verification item, we indicate the expected answer (Yes/No) and the classification of the possible defect identified according to the categorization presented in [19]: *omission*, *incorrect fact*, *ambiguity*, *inconsistency*, and *extraneous information*.

TABLE II. LGPDCHECK SUPPORTING MATERIAL FOR THE PRINCIPLE OF NECESSITY.

| LGPD Principle: Necessity |
|---|
| **1. Definition of the principle by the law:** limitation of the processing to the minimum necessary to achieve its purposes, covering data that are relevant, proportional, and non-excessive concerning the purposes of the data processing. |
| **2. Interpretation of the principle for software systems:** A software system should only collect personal data that is necessary and specific to achieve the functional purposes of the system.<br>There must be proportionality between the purpose of the designed functionality and the personal data collected to achieve that purpose. Therefore, when creating functionalities, organizations should focus on having only individual data sets in their domain to achieve the expected functionality. |
| **3. Action expected by the organization**: The data processing agent's collection of the user's data must consider the proportionality and the need for data collection to process personal data to reduce the amount of personal data collected, thus avoiding further abusive behavior. |
| **4. Example of violating the principle in software systems:** The principle of necessity is broken when a collection is based on "collecting as much as I can, just because I can," i.e., gathering and storing unnecessary data to fulfill the designed functionality.<br>Some business rules of a financial software system depend on the user's credit rating. While performing a credit assessment, running a query to an external system may be necessary to use personal identification data, such as CPF (Brazilian Tax Identification Number), address, and income data. In this scenario, it is inadequate to collect customer" data addressing sexual orientation or gender to carry out a credit assessment (purpose), which is irrelevant to the data procession activity. Besides violating the principle of necessity, this behavior also requires collecting sensitive data. |

| 5. Verification items | (5.A) Expected Answer | (5.B) Defect classification |
|---|---|---|
| (7) When processing data for the desired system functionality, are the data collection procedures limited to gathering only data explicitly required by the business rules to achieve the desired service or functionality (purpose)? | No | Extraneous information |
| (8) Is there any attempt to reduce or exclude any collected data from the data set to make it smaller? | No | Omission |

The *LGPDCheck* resources include an inspection report template adapted from [19]. In this report, the inspector should inform the inspected artifact, the inspection level applied, the description of each defect, its category, and its location.

IV. EXPERIMENTAL STUDY

The study design is mainly based on the study performed in [22] to characterize the feasibility of another checklist-based inspection technique. However, it is also based on the experience of the researchers developing and evaluating other inspection techniques (see Table I). We conducted a feasibility study to determine whether our proposed intervention (*LGPDCheck*) is appropriate for further testing. In this sense, we intend to assess whether or not the ideas and findings can be shaped to be relevant and sustainable. Besides, we want to identify the need (and how) to improve the research method and its protocols.

*A. Specific Goals*

Based on the GQM template [29], the goal of this feasibility study was defined as follows:

*Analyze*: the inspection of privacy and data protection features in software artifacts by using *ad-hoc* and *LGPDCheck* techniques

*For the Purpose of*: characterize

*With respect to*: the effectiveness (*defects identified/ total existing defects*) and efficiency (*identified defects/ time*) of *LGPDCheck* in detecting defects in the light of LGPD principles

*From the viewpoint of*: Software Engineering researchers

*In the context of*: software practitioners (represented by 3rd-year software engineering undergraduate students from UFRJ) developing IoT software systems.

*B. Questions and Metrics*

- *Q1:* How much time did the inspections take?
- *Metrics:* time dedicated to the inspection (in minutes), the efficiency of each inspection.
- *Q2:* Which inspection technique (*LGPDCheck* or *ad-hoc*) allows the inspectors to detect more defects?

- *Metrics:* number of defects detected, the effectiveness of each inspection.

The definitions of efficiency and effectiveness employed in this study follow the ones used in previous evaluations of other checklist-based software inspections [22,25,26,27,28]. By efficiency, we mean the study participant's average time (in minutes) to detect a single defect. By effectiveness, we mean the ratio between the number of defects found by the participant and the total number of distinct defects found by each participant

*C. .Hypotheses*

In this study, we will invite the participants to perform ad-hoc inspections (control) and inspections with the support of *LGPDCheck* (intervention). In this sense, we established the following hypotheses:

$H_01$: There is no difference between the efficiency of inspections conducted with *LGPDCheck* and *ad-hoc* inspections.
$H_A1$: The efficiency of inspections conducted with *LGPDCheck* differs from the efficiency of *ad-hoc* ones.
$H_02$: There is no difference between the effectiveness of the inspections conducted with *LGPDCheck* and *ad-hoc* inspections.
$H_A2$: The effectiveness of inspections conducted with *LGPDCheck* differs from that of ad-hoc ones.

*D. Variables*

- *Independent variables:* participant's inspection experience and previous knowledge of Privacy and Data Protection from regulations. All these variables will be collected through a subjects' characterization questionnaire, following the scale types presented in Table III.

TABLE III. INDEPENDENT VARIABLES

| Name (abbreviation) | Description | Scale Type |
|---|---|---|
| Experience with Software Development (SDExp) | Participant's previous experience with software development activities in general in terms of: self-evaluation/ years/ number of projects | 5-Level Likert scale (self)/Integer/ Integer |
| Experience with software inspection (SIExp) | Participant's previous experience in performing software inspections in terms of self-evaluation/ years/number of projects | 5-Level Likert scale (self) /Integer/ Integer |
| Knowledge about PDP (PDPK) | Participant's self-evaluation about their level of expertise about PDP | 5-Level Likert scale (self) |
| Experience with LGPD | Participant's previous experience in employing LGPD in software projects in terms of self-evaluation/ years/ number of projects | 5-Level Likert scale (self) /Integer/ Integer |

- *Dependent variables:* number of defects and false positives, time spent performing the inspection, efficiency, and effectiveness. The description of dependent variables is presented in Table IV.

TABLE IV. DEPENDENT VARIABLES DESCRIPTION

| Variable (abbreviation) | Description | Scale type |
|---|---|---|
| Number of Defects (ND) | Number of defects identified during the inspection by the participant | Numeric |
| False positives (FP) | Number of candidates to defect discarded | Numeric |
| Inspection duration (ID) | Time spent by the participant to perform the inspection (in minutes) | Numeric |
| Efficiency (Eff) | The average time (in minutes) to detect a single defect. | Numeric (Ratio) |
| Effectiveness (Eft) | The ratio between the number of defects found by the participant and the total number of distinct defects found by each participant | Numeric (Ratio) |

*E. Participants, Subjects, and Datasets*

This experimental study will be performed with 17 undergraduate students in Systems Engineering from the Federal University of Rio de Janeiro. These students work on a project named *Daily Huddle*[2], in which they are developing an IoT-based system designed to optimize communication and support the resource management of hospitals through identifying and allocating resources and tasks with specific responsibility assignments and deadlines. In this project, the 17 students are distributed between three groups (G1, G2, G3). Each group works to develop different system modules (M1, M2, and M3, respectively). Follows a brief description of each module:

- *M1* is a module for *managing hospital structural occurrences*, which includes occupancy of beds and operating conditions of beds (active beds and blocked beds).
- *M2* is a module for *process management*, which includes functionalities for monitoring the conditions of internal hospital procedures such as overload and urgencies and providing information for the professionals that oversee these procedures.
- *M3* is a module for team management, which includes shifts, personnel, and the management of occurrences with team members.

The three modules of the system will be inspected in the quasi-experiment. For this purpose, a balanced subset of functionalities consuming personal data will be selected to be inspected by the module. Thus, the criteria for balancing the scope of the inspections between modules will be the number of operations consuming personal data. No defect will be seeded. The authors will validate the defects reported by the participants. First, the first author will identify all the distinct defects reported by the participants for each module. Second, the authors will run a meeting to validate each one, composing the set of actual defects of each module. The following subsection (execution plan) describes the modules that the participants from each group will be manually assigned to inspect.

---

[2] https://status.net/templates/daily-huddle/

*F. .Execution Plan*

After signing in a consent form and a characterization form, each group of students (G1, G2, G3) will perform two trials (T1 and T2) over distinct combinations of modules (M1, M2, M3) of the software system they work (see Table V). The execution plan considers that the feasibility study involves the participation of the postgraduation students during their class time. Thus, all participants should receive the same instructions at each trial. The groups will only be assigned to inspect modules they had not teamed worked on during the development phase of the project/module.

The quasi-experiment will be taken in the classroom of the course, in which each participant has access to a single module. In the first trial (T1), all three groups will receive instructions about the LGPD and its principles. They will also receive instructions about software inspection and defect types. After the training, the participants will perform the ad-hoc inspections individually, reporting the discrepancies identified in a form.

TABLE V. GROUPS, TRIALS, AND INSPECTED ARTIFACTS

| Group (size) | T1 | Scope | T2 | Scope |
|---|---|---|---|---|
| G1 (5) | ad-hoc | M2 | *LGPDCheck* | M3 |
| G2 (6) | ad-hoc | M3 | *LGPDCheck* | M1 |
| G3 (6) | ad-hoc | M1 | *LGPDCheck* | M2 |

After T1, the participants will participate in the training session. In this session, they will be introduced to *LGPDCheck* and its resources. In the second trial (T2), the participants from each group will be assigned to perform inspections using *LGPDCheck* over other system modules they have yet to inspect but were already inspected ad-hoc by another group. As in the first round, the participants will report the discrepancies identified in a discrepancies report sheet. However, they should also associate each discrepancy with an item from the *LGPDCheck* checklist.

*G. Data Analysis*

The following steps will be followed for data analysis. First, we will identify the defects reported in each inspection and gather the time spent. Then, based on the set of distinct defects detected in each module, we will calculate the effectiveness of each inspection. Based on the manual analysis of participants' experience, we will distribute them between a group with more experience and a group with less experience. Next, we will compose distributions of effectiveness and efficiency grouped by trial (2), module/trial (6), and experience/trial (4). After eliminating outliers, we will verify the distributions' normality (Shapiro-Wilk) and homoscedasticity (Levene). If all verifications are well-succeeded, we apply the Student t parametric test to test the study hypotheses. If not, we will use the Mann-Whitney nonparametric test.

*H. Threats to Validity*

*Construct validity:* We highlight the small number of subjects and the limited number of inspected domains (one system, three modules), nonetheless, a limited combination of groups (three). However, it is worth noting that no participant is expected to inspect the same module more than once. Besides, we cannot guarantee that the modules (M1, M2, and M3) are comparable in the number of defects and complexity. To mitigate this threat, we composed a balanced subset of functionalities to inspect in each module.

*External validity:* The subjects are at least 3rd-year undergraduate students with different experience levels in software engineering and inspection.

*Internal validity:* LGPDCheck will be evaluated by the same groups of developers working in the software system to be inspected (modules M1, M2, and M3). Additionally, given the novelty and lack of materials focusing on PDP for software engineering, we are preparing training sessions so each participant will acquire knowledge on the subject.

*Conclusion Validity:* The initial results cannot be generalized. However, they will contribute to the evolution of LGPDCheck for further evaluation and use.

*I. Experimental Study Costs*

  *1) Planning Costs:*
   *a)* Plan itself: not applicable;
   *b)* Instrumentation: consent form, participants' characterization questionnaire, *LGPDCheck*, Inspection Report Template
   *c)* Training Material: video recording;
   *d)* Plan Evaluation: not applicable;

  *2) Execution Costs:*
   *a)* Travel expenses: local application;
   *b)* Training: Training on the main concepts of Privacy and Data Protection from LGPD and training on LGPDCheck will be provided to students;
   *c)* Human Resources: undergraduate students and researchers;
   *d)* Material Resources: forms, software artifacts, computers, software.

  *3) Analysis Costs:* not applicable;
  *4) Packaging Costs:* not applicable;
  *5) Reporting and Publishing:* conference registration and open access fees.

## V. CONCLUSION AND FUTURE WORK

We believe that our research will contribute to advancing the field of privacy engineering and support organizations in complying with privacy regulations. By adopting a proactive approach to privacy protection, we can ensure that personal data is safeguarded and used ethically in our digital society. In particular, the Brazilian software industry has an urgency in assuring the compliance of their systems with the LGPD. After an adaptation period, the LGPD is being applied in the country.

Therefore, we intend to replicate this study by applying LGPD in external projects if we achieve satisfactory results in this execution. In the replications, we will evaluate the technique's efficiency and effectiveness by users and stakeholders in software projects outside our portfolio. Considering that LGPD just started to be effectively applied in Brazil, we expect to identify more specialists to contribute to future evaluations. Besides, we will evaluate the acceptance of the proposed technology among the study participants through an opinion questionnaire. If the results are positive, we will promote *LGPDCheck* as a recommended technique for privacy protection in software systems artifacts. Otherwise, we will use the findings to identify the deficiencies of the technique and work on evolving it.

ACKNOWLEDGMENTS

CAPES, CNPq, and FAPERJ support this research in the context of the Engineering of Contemporary Software Systems in Brazil.